\documentclass[
    aps,
    nobibnotes,
    reprint,
    prl,
    letterpaper,
    nofootinbib,
    amsfonts,
    amssymb,
    amsmath,
    superscriptaddress
]{revtex4-2}

\usepackage{mathtools}
\usepackage{ragged2e}
\usepackage{color}
\usepackage{bm}
\usepackage[version=4]{mhchem}
\usepackage{hyperref}
\usepackage{xspace}
\usepackage{soul,xcolor}\setstcolor{red}

\newcommand{\rucl}{$\alpha$-RuCl$_3$\xspace}

\begin{document}

\title{Sliding Disassembly of van der Waals Heterostructures}

\author{Jordan Pack}
\thanks{Equal contribution}
\affiliation{Department of Physics, Columbia University, New York, NY, USA}

\author{Karl V. Falb}
\thanks{Equal contribution}
\affiliation{Department of Applied Physics and Applied Mathematics, Columbia University, New York, NY, USA}

\author{Sanat Ghosh}
\affiliation{Department of Physics, Columbia University, New York, NY, USA}

\author{Xuehao Wu}
\affiliation{Department of Physics, Columbia University, New York, NY, USA}

\author{Keng Tou Chu}
\affiliation{Department of Physics, University of Washington, Seattle, WA, USA}

\author{Florie Mesple}
\affiliation{Department of Physics, University of Washington, Seattle, WA, USA}

\author{Ellis Thompson}
\affiliation{Department of Physics, University of Washington, Seattle, WA, USA}

\author{Zhuquan Zhang}
\affiliation{Department of Physics, Columbia University, New York, NY, USA}

\author{Carolin Gold}
\affiliation{Department of Physics, Columbia University, New York, NY, USA}

\author{Kenji Watanabe}
\affiliation{Research Center for Electronic and Optical Materials, National Institute of Materials Science, 1-1 Namiki, Tsukuba, Japan}

\author{Takashi Taniguchi}
\affiliation{Research Center for Materials Nanoarchitectonics, National Institute of Materials Science, 1-1 Namiki, Tsukuba, Japan}

\author{Dmitri N. Basov}
\affiliation{Department of Physics, Columbia University, New York, NY, USA}

\author{A. N. Pasupathy}
\affiliation{Department of Physics, Columbia University, New York, NY, USA}
\affiliation{Condensed Matter Physics and Materials Science Division, Brookhaven National Laboratory, Upton, NY, USA}

\author{Matthew Yankowitz}
\affiliation{Department of Physics, University of Washington, Seattle, WA, USA}
\affiliation{Department of Materials Science and Engineering, University of Washington, Seattle, WA, USA}

\author{Cory R. Dean}
\email{cd2478@columbia.edu}
\affiliation{Department of Physics, Columbia University, New York, NY, USA}

\author{Aravind Devarakonda}
\email{aravind.devarakonda@columbia.edu}
\affiliation{Department of Applied Physics and Applied Mathematics, Columbia University, New York, NY, USA}

\maketitle

\clearpage

\setlength{\parindent}{4ex}
\noindent
\justify

\label{Abstract}

\textbf{\noindent Many recent advances in our understanding of two-dimensional (2D) electron systems stem from van der Waals (vdW) heterostructures. The assembly process relies on the weak bonding across interfaces between layered vdW compounds, making it possible to construct exceptionally clean heterostructures from chemically and structurally distinct materials--a challenging task for traditional thin-film growth techniques. Here we demonstrate an additional, dynamic degree of freedom afforded by vdW interfaces, wherein we use microstructured polymer stamps to disassemble and reconfigure vdW heterostructures by sliding. We apply this technique to alter the dielectric environment of monolayer graphene, perform scanning tunneling microscopy on semiconducting and air-sensitive monolayers, and manipulate strain-sensitive moir\'e materials. Together these demonstrations suggest a new paradigm for assembling and dynamically modifying van der Waals heterostructures, with the potential to reveal new insights into 2D electron systems. }


\begin{figure*}
    \makebox[\textwidth][c]{\includegraphics[scale=1.0]{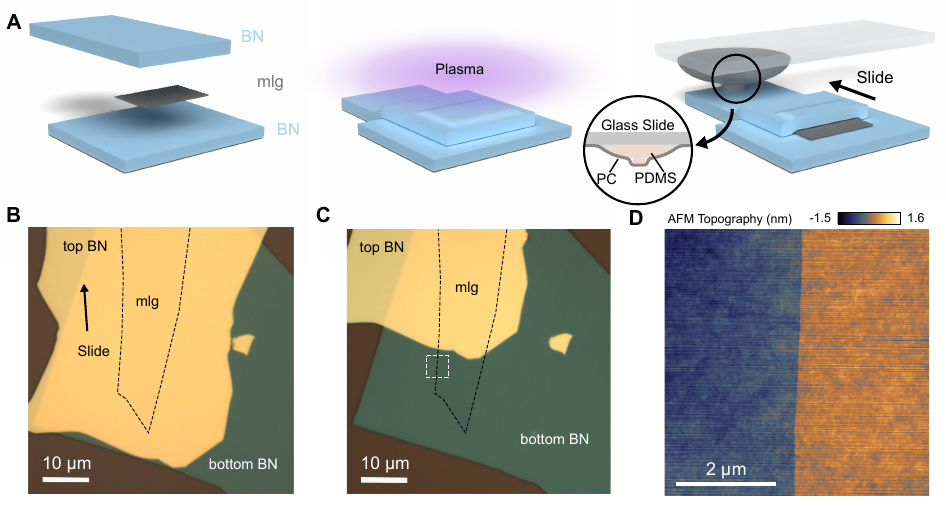}}
\caption{\textbf{Sliding disassembly of van der Waals heterostructures.} \textbf{(A)} Schematic illustration of the sliding disassembly technique. First, a heterostructure is assembled using the conventional dry-assembly technique, followed by a short plasma treatment to clean the surface. To manipulate the layers, we use a PDMS stamp designed to limit the contact area to the target layer for sliding. By contacting the top layer with the sample and sliding laterally, for example, it can be removed from the heterostructure to reveal the layer underneath. \textbf{(B)} Optical image of an encapsulated monolayer graphene (mlg) before sliding the top hBN. \textbf{(C)} Optical image of the device in \textbf{(B)} after sliding the top BN to partially reveal the monolayer graphene. \textbf{(D)} AFM topograph of the exposed graphene edge (white box in \textbf{(C)}), revealing a clean, flat monolayer graphene surface.}
\end{figure*}

\begin{figure*}
    \makebox[\textwidth][c]{\includegraphics[scale=1.0]{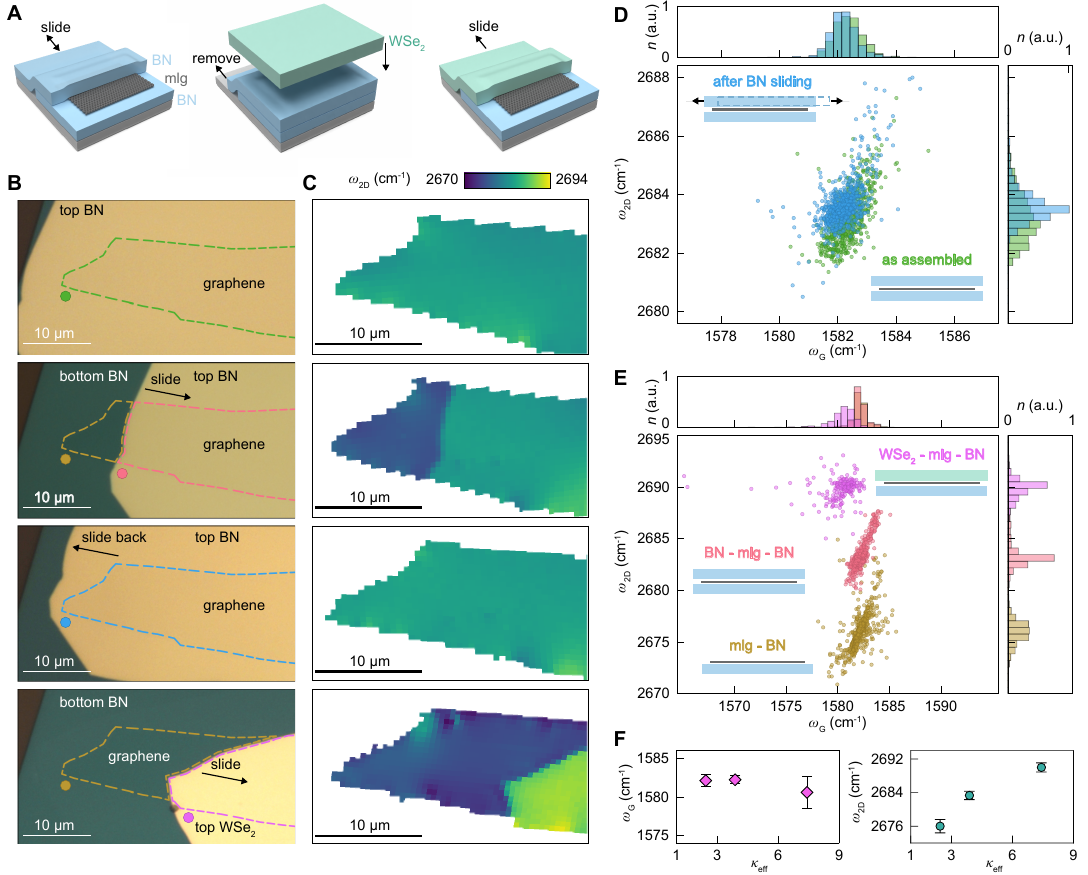}}
\caption{\textbf{Reconfigurable van der Waals heterostructures.} \textbf{(A)} Schematic reconfiguration of a vdW heterostructure by replacing the top encapsulating material. \textbf{(B)} Optical images shown at multiple stages of the reconfiguration. After standard dry-transfer assembly, the top hBN is shifted to partially reveal the underlying monolayer graphene, then shifted back to cover it. Next the top hBN is completely removed, exfoliated 2$H$-WSe$_2$ is placed over top, and then shifted to partially expose the graphene. \textbf{(C)} The spatial variation of the graphene 2D Raman peak $\omega_\mathrm{2D}$ for each stage pictured in \textbf{(B)}. \textbf{(D)} Scatter plot of the graphene Raman G ($\omega_\mathrm{G}$) and 2D ($\omega_\mathrm{2D})$ peaks of fully hBN-encapsulated graphene as assembled and after sliding manipulation, showing minimal variation. \textbf{(E)} Scatter plot of $\omega_\mathrm{G}$ and $\omega_\mathrm{2D}$ in the partially exposed stages, illustrating the systematic shift in $\omega_{2D}$ due to differences in the dielectric environment surrounding the graphene sheet. \textbf{(F)} Median values of (left) $\omega_\mathrm{G}$ and (right) $\omega_\mathrm{2D}$ histograms, with one standard deviation error bars, as a function of the effective dielectric constant $\kappa_\mathrm{eff}$.}
\end{figure*}

\begin{figure*}
    \makebox[\textwidth][c]{\includegraphics[scale=1.0]{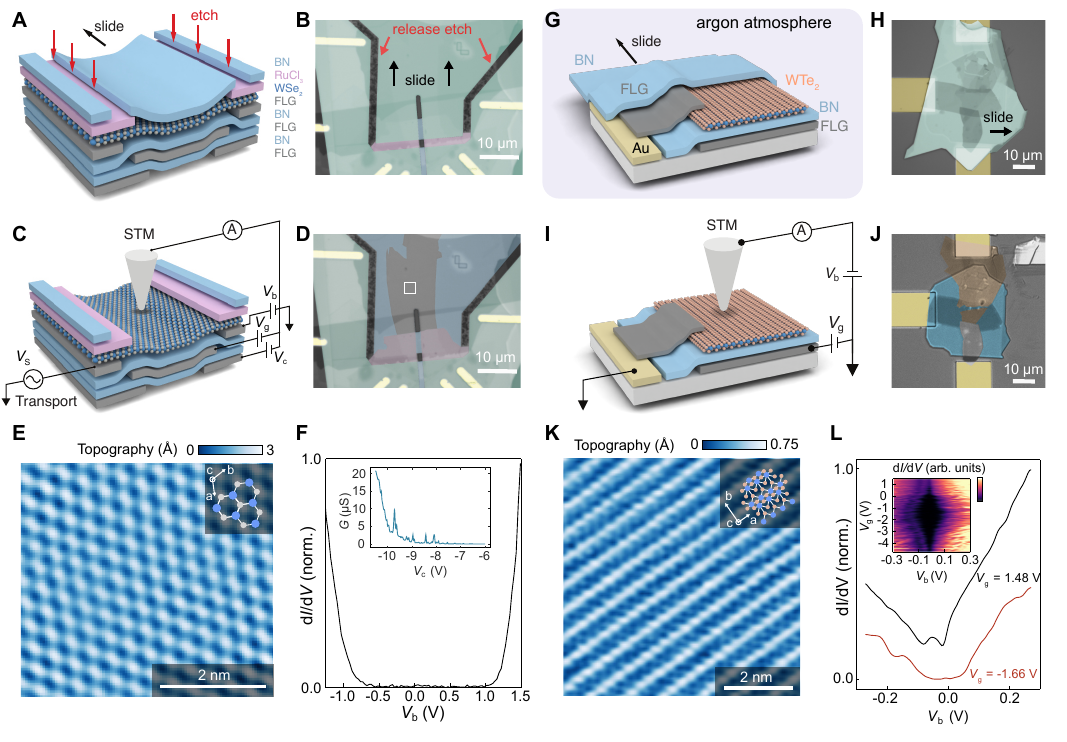}}
\caption{\textbf{Scanning tunneling microscopy of disassembled devices.} \textbf{(A)} Schematic cartoon of the assembled monolayer $H$-\ce{WSe2} heterostructure with charge-transfer contacts, channel gate, and contact gate. The top hBN layer is etched into a removable region using an etch procedure designed to stop on the \rucl layer over the contacts.  After etching, this region is slid to expose the device channel for \textbf{(C)} STM measurements. False color image of the device \textbf{(B)} before and \textbf{(D)} after the sliding release step. \textbf{(E)} Scanning atomic resolution topography of a gate-tuned region of the sample. (inset) Overlaid $H$-\ce{WSe2} crystal structure with tungsten (blue) and selenium (gray) atoms. \textbf{(F)} $dI/dV$ spectroscopy of the sample showing a 2.11 $\pm$ 0.08 eV band gap and Fermi level offset due to tip induced band bending. (inset) Two-terminal conductance of the sample as a function of contact gate measured at low temperature prior to removing the top hBN. \textbf{(G)} Schematic illustration of the sample consisting of a sacrificial top hBN, monolayer 1$T'$-\ce{WTe2}, bottom hBN, with a graphite contact and back gate. \textbf{(H)} False colored optical micrograph of the heterostructure as assembled. \textbf{(I)} Schematic of the final device heterostructure. \textbf{(J)} False color image of device after removing the top hBN layer inside a glovebox. \textbf{(K)} STM topograph of monolayer 1$T'$-\ce{WTe2}. (inset) Overlaid 1$T'$-\ce{WTe2} crystal structure with tungsten (blue) and tellurium (yellow) atoms. \textbf{(L)} $dI/dV$ spectrum measured at two different values of $V_\mathrm{g}$ averaged within $V_\mathrm{g}$ = 0.04 V windows. (inset) $dI/dV$ spectroscopy map versus $V_\mathrm{g}$.}
\end{figure*}

\begin{figure*}
    \makebox[\textwidth][c]{\includegraphics[scale=1.0]{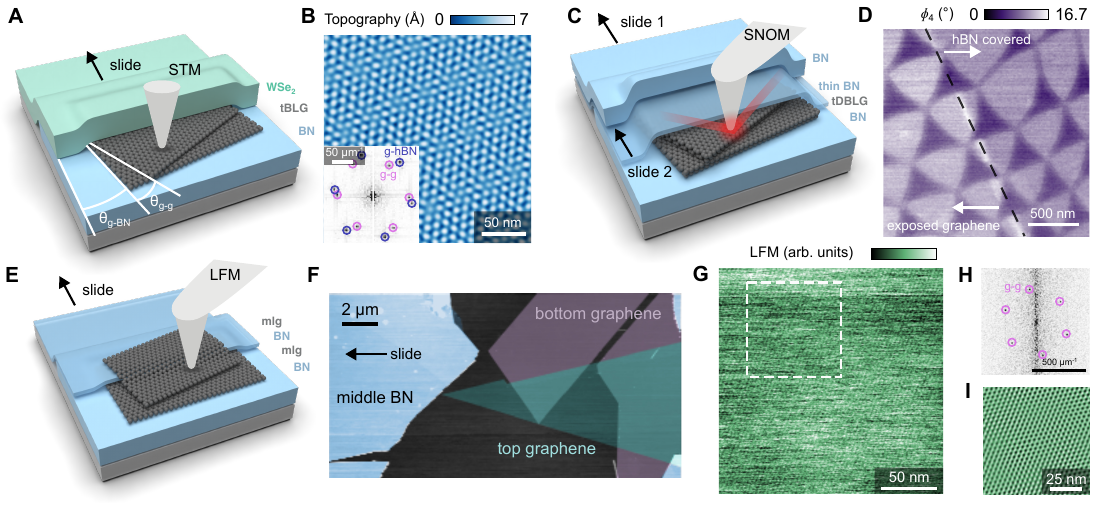}}
\caption{\textbf{Sliding disassembly of moir\'e heterostructures.} \textbf{(A)} Schematic illustration of twisted bilayer graphene (tBLG) aligned to hBN for STM characterization, with graphene-graphene twist angle $\theta_\mathrm{g-g}$ and graphene-hBN twist angle $\theta_\mathrm{g-BN}$. In this device, exfoliated 2$H$-\ce{WSe2} is used as the removable top layer. \textbf{(B)} STM topograph of the hBN-aligned twisted graphene device. The inset Fourier transform evidences two coexisting periodicities – one from the twisted graphene moir\'e and one from the graphene-hBN moir\'e. \textbf{(C)} Schematic of a twisted double bilayer graphene (tDBLG) sample measured with scanning near-field optical microscopy (SNOM). Two pulling steps are used – first to remove a topmost hBN layer to reveal a clean thin hBN surface, and a second to move the thin hBN partially to reveal the tDBLG. \textbf{(D)} Mid-infrared near-field phase image traversing the thin hBN edge showing domains of rhombohedral (dark) and Bernal (light) stacking orders. A dashed line indicates the edge of the thin hBN layer. The underlying domain structure evolves continuously across this boundary. \textbf{(E)} Schematic cartoon for the construction of a tBLG device by extracting a middle hBN spacer that initially separates the two graphene layers. \textbf{(F)} False-color AFM topography image of a tBLG prepared by technique. \textbf{(G)} Lateral force microscopy (LFM) image of the tBLG moir\'e formed by extracting the middle hBN of the device shown in \textbf{(F)}. \textbf{(H)} Fast Fourier transform of \textbf{(G)}. \textbf{(I)} Fourier filtered section of \textbf{(G)} (white dashed window) highlighting the moir\'e superlattice.}

\end{figure*}


Heterostructures fabricated from two-dimensional (2D) materials are rich and highly tunable platforms for studying novel electronic states. With strong in-plane bonding and weaker van der Waals (vdW) bonds oriented out-of-plane, 2D materials may be layered with little concern for interlayer chemistry and lattice registry. This makes it possible to interface materials with wide-ranging properties, free from the chemical considerations that constrain conventional heterostructure synthesis \cite{geimVanWaalsHeterostructures2013}. The unique, additional degrees of freedom afforded by vdW interfaces are exemplified by moir\'e patterning, where lattice mismatch and rotational faults at 2D interfaces can be exploited to engineer band structures supporting new correlated and topological ground states \cite{balentsSuperconductivityStrongCorrelations2020, kennesMoireHeterostructuresCondensedmatter2021}.

Experimental progress with vdW heterostructures has stemmed from continued advances in assembly methods. For example, mechanical transfer, in which exfoliated layers are physically laminated together, has enabled the fabrication of increasingly complex geometries with pristine 2D interfaces \cite{deanBoronNitrideSubstrates2010,wangOneDimensionalElectricalContact2013,kimVanWaalsHeterostructures2016,saitoIndependentSuperconductorsCorrelated2020,liuDisassembling2DVan2020, mannixRoboticFourdimensionalPixel2022,wangCleanAssemblyVan2023}. In tandem with heterostructure construction, various electrical contact methods have been developed to probe individual layers within multilayer and multi-gated architectures \cite{wangOneDimensionalElectricalContact2013,liExcitonicSuperfluidPhase2017, telfordMethodLithographyFree2018,zengHighQualityMagnetotransportGraphene2019, packChargetransferContactsMeasurement2024}. More recently, dynamically reconfigurable device structures have emerged that allow \textit{in-situ} modification of rotational and translational alignment between layers \cite{ribeiro-palauTwistableElectronicsDynamically2018,kapferProgrammingTwistAngle2023,zhangDynamicTwistingImaging2023,wakafuji3DManipulation2D2020,barabasMechanicallyReconfigurableVan2023}. 

The predominant focus throughout these efforts has been to maximize control while minimizing disorder during the heterostructure assembly. Despite these technical advances, we still lack a robust technique to controllably disassemble these heterostructures in order to gain access to or manipulate subsurface layers.  Existing techniques rely either on the destructive removal of capping layers through chemical etching \cite{agarwalSituEngineeringHexagonal2025,nipaneDamageFreeAtomicLayer2021}, or specialized processes that result in ``open face'' devices during assembly \cite{kerelskyMaximizedElectronInteractions2019,liuVisualizingBrokenSymmetry2022,choiElectronicCorrelationsTwisted2019,kimImagingIntervalleyCoherent2023,thompsonMicroscopicSignaturesTopology2025, taoReconfigurableElectronicsDisassembling2021}. In all of these cases, the accessible 2D materials and device geometries are constrained, and process-induced disorder remains a primary concern.

Here we introduce a novel, dynamic method for disassembling and reconfiguring vdW heterostructures that is non-destructive, reversible, and applicable at any step in the fabrication process including the layer lamination step, lithographic processing stage, or even after the device is fully processed and electrically connected for measurement. Our method takes advantage of the low interfacial friction between the vdW layers \cite{liaoUItralowFrictionEdgepinning2022} by using a microstructured, molded polymer stamp to achieve deterministic removal of targeted layers from the heterostructure by controlled sliding. Scanning probe microscopy, Raman spectroscopy and moir\'e imaging all suggest the sliding procedure does not appreciably deform or otherwise alter the constituent layers. Our technique works with high repeatability on all combinations of vdW materials that we have attempted, including both hetero- and homo-interfaces. Moreover, the sliding can be used to remove, return, or replace surface capping layers as well as intermediate layers encapsulated within the vdW heterostructure. We demonstrate the utility of this new capability in several ways, including reconfiguring the dielectric environment of an assembled heterostructure, fabricating scanning probe compatible devices by disassembling heterostructures containing moir\'e patterned, air-sensitive, and semiconductor 2D materials, and creating moir\'e patterns between rotationally faulted encapsulated layers through a novel lamination process.

\section{Sliding disassembly}

\noindent The sliding disassembly method is demonstrated in Fig. 1, using hexagonal boron nitride (hBN) encapsulated monolayer graphene as an illustrative example. Figure 1A shows a schematic overview of the process. The encapsulated heterostructure is first assembled using the conventional dry transfer technique \cite{wangOneDimensionalElectricalContact2013}, followed by solvent cleaning and brief plasma treatment. Separately, a microstructured PDMS stamp consisting of a dome terminated at its apex by a cylindrical micropillar is prepared on a glass microscope slide, and then draped with a polycarbonate (PC) film (Fig. SI 1). We use a number of micropillar geometries, with diameters between 40 to 100 $\mu$m and heights in the range of 20 to 50 $\mu$m. Micropillars of these sizes provide a focused contact area with the target layer, in this case hBN, and the PC film allows temperature controlled adhesion (see Methods). To perform disassembly, the PC covered micropillar is brought into contact with the top hBN layer and then translated to initiate sliding. For a fully encapsulated layer, we find that translation of the top layer does not cause any movement of the lower layers, whereas translation of the bottom layer tends to move the entire heterostructure (Fig. SI 2).

Figures 1B and 1C show optical images of a device before and after sliding the top hBN, respectively, to partially reveal monolayer graphene. Atomic force microscopy (Fig. 1D) confirms that the graphene is atomically flat, as expected for graphene on hBN substrates \cite{deanBoronNitrideSubstrates2010}, and residue free. It is noteworthy that the graphene region exposed by sliding has never been in contact with polymer, and thus eliminates the need for additional chemical or thermal cleaning procedures \cite{nuckollsQuantumTexturesManybody2023}.

We have successfully applied this same procedure to more than 30 heterostructures, with constituent layers comprising mono- and multilayer graphene, hBN, transition metal dichalcogenide (TMDs), and mica. We have observed an estimated 75\% success rate across all material combinations, defined as successfully sliding the target layer the intended distance without functionally disturbing the remaining heterostructure. In the 25\% of unsuccessful attempts, we note three dominant failure modes: (1) the target layer does not slide, or instead the entire heterostructure slides together. We attribute this to  unintentional alignment between nearly lattice-matched layers that results in increased interlayer friction \cite{ribeiro-palauTwistableElectronicsDynamically2018}. (2) The sliding layer fractures. This generally correlates with visible defects and/or localized pinning of one part of the sliding layer by residue or edges. Tearing is observed with higher frequency when attempting to slide targets that are only a few atomic layers thick. (3) The stationary layer tears. In nearly all cases where this happens, we can identify a visible defect that was unintentionally pulled across the material causing the tear.

\section{Reconfiguring heterostructures}

\noindent We utilized Raman spectroscopy to interrogate whether the mechanical sliding shown in Fig. 1 imparts measurable strain to the encapsulated graphene layer. We performed the measurements in sequential stages, as illustrated in Fig. 2A. First we assembled a an hBN – monolayer graphene – hBN heterostructure, and then partially slid off the top hBN to reveal a portion of the encapsulated graphene. We then reversed the process by pushing the top hBN back into place, re-encapsulating the graphene. Finally, we fully removed the top hBN by sliding it off the heterostructure, replaced it with an exfoliated flake of 2$H$-\ce{WSe2} using conventional transfer methods, and then partially slid it to re-expose a region of the graphene. A video of this sequence is available as Supplemental Video 1.

Optical images (Fig. 2B) and spatial maps of the graphene Raman 2D peak position $\omega_\mathrm{2D}$ (Fig. 2C) are shown for select stages of the manipulation sequence: as-fabricated with graphene fully encapsulated, top hBN partially retracted, top hBN replaced, and top 2$H$-\ce{WSe2} partially retracted. Figure 2D shows a scatter plot and associated histograms of $\omega_\mathrm{2D}$ versus the graphene Raman G peak position $\omega_\mathrm{G}$ for hBN covered graphene as fabricated (green) and then after retracting and repositioning the top hBN (blue). Each point corresponds to a single pixel of the associated map in Fig. 2C. 

Lattice strain in graphene monolayers can be identified from the $\omega_\mathrm{2D}$ and $\omega_\mathrm{G}$ peak shifts \cite{leeOpticalSeparationMechanical2012}. Before sliding, we measure $\omega_\mathrm{2D} = 2683.2$  $\pm$ 0.9 cm$^{-1}$ and $\omega_\mathrm{G} = 1582.4$ $\pm$ 0.5 cm$^{-1}$. After sliding and re-encapsulating, we find $\omega_\mathrm{2D} = 2683.5$ $\pm$ 0.8 cm$^{-1}$ and $\omega_\mathrm{G} = 1582.2$ $\pm$ 0.5 cm$^{-1}$ which, within error, indicates $\omega_\mathrm{2D}$ and $\omega_\mathrm{G}$ do not change across this sliding sequence, and that it is a non-destructive, reversible process. From the minimal change in the mean values of $\omega_\mathrm{2D}$ and $\omega_\mathrm{G}$, we can place an upper bound of -0.02\% strain imparted during the sliding process, which is an order of magnitude smaller than the strain deduced from scanning tunneling microscopy (STM) studies of moir\'e heterostructures assembled with conventional dry transfer methods \cite{mespleHeterostrainDeterminesFlat2021}.

Figure 2E compares $\omega_\mathrm{2D}$ versus $\omega_\mathrm{G}$ peak positions for three different top layer configurations; open to air (yellow), hBN covered (red), and 2$H$-\ce{WSe2} covered (purple). In this comparison, we see a systematic shift in $\omega_\mathrm{2D}$ with minimal change in $\omega_\mathrm{G}$. The reversibility demonstrated in Fig. 2D suggests that the observed shift is not a consequence of the mechanical manipulation but rather due to  differences in the dielectric environment surrounding the monolayer graphene. Figure 2F plots the mean $\omega_\mathrm{2D}$ (green) and $\omega_\mathrm{G}$ (purple) positions, obtained from the histograms in Fig. 2E, versus the inferred, effective dielectric constant $\kappa_\mathrm{eff}$ surrounding the monolayer graphene (Methods). We see from Fig. 2F that within error $\omega_\mathrm{G}$ is independent of  $\kappa_\mathrm{eff}$, whereas $\omega_\mathrm{2D}$ increases monotonically. This observation is consistent with calculations that suggest electronic screening of the phonon dispersion can modify $\omega_\mathrm{2D}$ while leaving $\omega_\mathrm{G}$ unchanged \cite{forsterDielectricScreeningKohn2013,lazzeriImpactElectronelectronCorrelation2008}. Further studies could reveal the detailed functional relationship between $\kappa_\mathrm{eff}$ and $\omega_\mathrm{2D}$ and fundamental insights into electron-phonon interactions at 2D vdW interfaces. The sliding disassembly process is invaluable towards this effort and other systematic studies of proximity effects and environment engineering at 2D interfaces, largely eliminating the confounding effects of device-to-device variation.

\section{Disassembling devices for scanning probe measurements}

\noindent Next, we demonstrate the utility of the sliding disassembly technique for preparing scanned probe-compatible device geometries.
Low temperature STM of 2D semiconductors poses a number of challenges, including stringent constraints on surface cleanliness, contact transparency, and complicating tip-gating effects \cite{floresVoltageDropExperiments1984,feenstraElectrostaticPotentialHyperbolic2003}. Sophisticated methods have been developed to tackle each of these issues \cite{liImagingMoireFlat2021,liImagingTwodimensionalGeneralized2021,thompsonMicroscopicSignaturesTopology2025,liuVisualizingBrokenSymmetry2022,xiangImagingQuantumMelting2025}, but addressing them simultaneously compounds the difficulty. Figures 3A through 3F show the first scanning tunneling microscopy (STM) measurements of gated monolayer $H$-\ce{WSe2}. In our approach, we begin by constructing a fully-encapsulated $H$-\ce{WSe2} transport device based on \rucl doped contacts that incorporates both a channel gate ($V_\mathrm{g}$) and contact gate ($V_\mathrm{c}$) to achieve low contact resistance to the low-density $H$-\ce{WSe2} channel \cite{packChargetransferContactsMeasurement2024}. The multilayer structure is assembled using the conventional dry transfer method, followed by etching and contacting with standard lithographic techniques (the schematic structure and optical image of a fabricated device are shown in Fig. 3A and 3B, respectively). To prepare this sample for subsequent STM characterization, we plasma etch a ``release'' cut through the topmost hBN layer to create an isolated section that can later be slid away to expose the underlying $H$-\ce{WSe2} layer. The \rucl layer serves as an etch stop, preventing the plasma from cutting through the contact region underneath. No further cleaning steps are performed after the hBN is removed.

Figures 3C and 3D show a schematic illustration and optical image of the device, respectively, after removing the top hBN, and Figure 3E shows an STM topograph acquired at the approximate location indicated by the white box in Fig. 3D (see methods for bias configuration during scanning). The fact that we achieve atomic resolution imaging confirms that the exposed surface is pristine and free of debris. 

Figure 3F shows a scanning tunneling spectroscopy (STS) curve acquired by varying the tip-sample bias $V_\mathrm{b}$ while maintaining a fixed voltage difference to the contact and channel gates (Methods), wherein we observe a band gap of 2.11 eV as expected for a $H$-\ce{WSe2} monolayer \cite{yankowitzLocalSpectroscopicCharacterization2015,zhangProbingCriticalPoint2015}. Previous measurements required the TMD to be on a conductive substrate to avoid complicating lateral voltage drops in the semiconducting material or at the metal-semiconductor interface, but by ensuring low contact and channel resistance, it is now possible to perform accurate spectroscopy in a gate-tunable device. Additionally, this device architecture makes it possible to measure electronic transport in the same sample as a complementary probe of the electronic properties. Figure 3F inset shows two terminal conductance as a function of $V_\mathrm{c}$ at $V_\mathrm{g}=-6$ V. At the gate voltage used, STS indicates the Fermi level lies in the band gap, while transport suggests the sample is doped to the valence band. This discrepancy arises from band bending due to the work function mismatch between the $H$-\ce{WSe2} monolayer and the tungsten STM tip, which we are able to identify by performing transport and STS on the same device. More broadly, our ability to perform both transport and STS in this single device highlights how sliding disassembly allows heterostructure optimization for multiple measurement modalities, in this case permitting complex contact engineering and the preparation of the pristine surface necessary for atomic resolution STM and STS.

Next, we used the disassembly technique to study monolayer 1$T'$-\ce{WTe2} by STM and STS. Due to its air sensitivity, STM measurements on this material have largely been performed on films grown by molecular beam epitaxy that are transferred through ultra-high vacuum (UHV) into a connected STM. While this avoids air exposure, this approach is technically challenging, limits substrate choice, and thus far with 1$T'$-\ce{WTe2} prevents gate tunability \cite{tangQuantumSpinHall2017,jiaDirectVisualizationTwodimensional2017,maximenkoNanoscaleStudiesElectric2022}. In parallel, efforts have been undertaken to fabricate gate-tunable STM devices from 1$T'$-\ce{WTe2} in inert glovebox environments that are protected with monolayer hBN to allow for solvent, mechanical, and thermal cleaning. While these samples can be probed by STM, the monolayer hBN overlayer introduces an additional, generally undesirable, tunneling barrier \cite{jiaEvidenceMonolayerExcitonic2022}.

In our approach, we start by assembling an encapsulated monolayer 1$T'$-\ce{WTe2} device inside an argon-filled glovebox by standard dry transfer. In sequence, we used a thick hBN flake to pick up graphite as the electrical contact, a monolayer of 1$T'$-\ce{WTe2}, a hBN bottom dielectric, and a bottom graphite gate. The vdW heterostructure was then deposited onto an Si/\ce{SiO2} wafer with predefined metal leads (Fig. 3G and 3H). Being fully encapsulated in the argon glovebox, we were able to move the device into ambient conditions to mount and electrically connect it to the STM sample holder without degradation of the 1$T'$-\ce{WTe2} layer. Once fully bonded to the STM platform, we returned the sample to the argon glovebox, removed the top hBN to expose the pristine monolayer of 1$T'$-\ce{WTe2}, and transferred it into the UHV STM chamber using an inert-gas suitcase, all within an hour. Figures 3I and 3J show a schematic illustration and optical microscope image of the device, respectively.

Figure 3K shows an STM topograph of monolayer 1$T'$-\ce{WTe2} with an overlay of the in-plane crystal structure. The atomic lattice is well-resolved and a larger area topograph shows no tip changes or scanning artifacts (Fig. SI 4A), indicative of a clean surface with negligible degradation. Figure 3L shows representative differential conductance d$I$/d$V$ versus tip bias $V_\mathrm{b}$ line cuts measured by STS at two different gate voltages $V_\mathrm{g} = -1.66$ V and $V_\mathrm{g} = 1.48$ V, corresponding to the sample being at charge neutrality and electron-doped, respectively. In both of these traces, we see a dip in d$I$/d$V$ at $V_\mathrm{b} = 0$ that is wider when the monolayer 1$T'$-\ce{WTe2} is tuned to charge neutrality. This widening of the zero-bias dip in the insulating regime near charge neutrality is more clearly visible in the d$I$/d$V$ map versus $V_\mathrm{b}$ and $V_\mathrm{g}$ (Fig. 3L, inset). In prior STM results from monolayer 1$T'$-\ce{WTe2} protected by monolayer hBN, this bias-dependent widening of the gap was interpreted as the emergence of a strongly-correlated, many-body state \cite{jiaEvidenceMonolayerExcitonic2022}. Using the same sliding methodology, we also examined a second device with monolayer 1$T'$-\ce{WTe2} resting directly on few-layer graphite (Fig. SI 4B), and reproduce features observed in MBE grown monolayer 1$T'$-\ce{WTe2} on conducting substrates. All together, the characteristics we observe are consistent with earlier results, but obtained in a more versatile geometry that makes further characterization of these phenomena simpler \cite{tangQuantumSpinHall2017,jiaDirectVisualizationTwodimensional2017,maximenkoNanoscaleStudiesElectric2022,jiaEvidenceMonolayerExcitonic2022,queGateTunableAmbipolarQuantum2024}. 

\section{Compatibility with moir\'e materials}

\noindent The sliding disassembly process is also compatible with moir\'e heterostructures. Figure 4A and 4B show STM  measurements of a twisted bilayer graphene (TBG) sample on hBN.  This device was prepared by using thick 2$H$-\ce{WSe2} to assemble the graphene and hBN using using the conventional cut and stack method \cite{saitoIndependentSuperconductorsCorrelated2020}. After lithographically defining contacts to the graphene, the 2$H$-\ce{WSe2} was then removed by sliding (Fig. 4A). Figure 4B shows atomically resolved topographic imaging. The fast Fourier transform (FFT) (Fig. 4B, inset) evidences two periodic structures -- one arising from the TBG interface and the other from lattice alignment between the hBN substrate and graphene layers. Analyzing the anisotropy of the moir\'e periodicities, we estimate uniaxial strain reaching 0.2\%, comparable to the typical strain found in conventional devices \cite{mespleHeterostrainDeterminesFlat2021}. This suggests that sliding did not impart significant additional strain to the moir\'e pattern.

To further characterize the impact of sliding on moir\'e patterns, we used scanning near-field optical microscopy (SNOM) to map a partially exposed moir\'e pattern formed by twisted double bilayer graphene (tDBLG). At small twist angles, tDBLG forms large area domains of rhombohedral and Bernal stacking that are distinguishable by SNOM \cite{halbertalMoireMetrologyEnergy2021} and, moreover, the domain shapes are particularly sensitive to local strain, allowing us to precisely map strain variation across moir\'e patterned devices \cite{yankowitzElectricFieldControl2014,liGlobalControlStackingOrder2020,halbertalMoireMetrologyEnergy2021,hsiehDomainDependentSurfaceAdhesion2023}.

The heterostructure we examine consists of a thick top hBN that we used to pick up in sequence a layer of thin hBN ($<$ 2 nm), tDBLG, and a bottom hBN. After releasing onto an Si/\ce{SiO2} substrate, we fully removed the topmost hBN layer to reveal the thin hBN surface. The result is a residue free and transparent hBN capping layer through which we can perform SNOM. We then partially retracted the thin capping layer to allow direct comparison of the moire pattern between the exposed and capped regions (Fig. 4C). Figure 4D shows a SNOM image across the hBN edge. The moir\'e domains evolve continuously and smoothly across the boundary, free of pinning or strain gradients. These observations further reinforce that sliding does not alter underlying moir\'e patterns.

Finally, we show that sliding enables a new method for assembling moir\'e heterostructures. Figure 4E illustrates a novel lamination technique to construct twisted bilayer graphene, wherein two rotationally faulted graphene monolayers are stacked with a hBN spacer partially separating them. We then remove the hBN spacer layer by pulling it out from between the top and bottom graphene layers -- laminating them together in the process. This local lamination method contrasts conventional moir\'e assembly where lamination proceeds by sequential pickup of exfoliated graphene monolayers from a Si/\ce{SiO2} substrate \cite{kimVanWaalsHeterostructures2016}. Figure 4F shows a false colored AFM topograph of a TBG device assembled through this approach. The overlap area is free of bubbles or wrinkles, and lateral force microscopy (LFM) in the TBG region reveals a moir\'e superlattice (Fig. 4G). From the FFT of the LFM image (Fig. 4H) we extract a twist angle of $\theta = 3.6^\circ$, and based on the sharpness of the FFT peaks, conclude that the superlattice is highly uniform over the scan window.

\section{Discussion}

\noindent The sliding control that we demonstrate provides a new degree of freedom that expands the possibilities for vdW heterostructure fabrication. Concretely, the low friction makes it possible to slide layers deterministically without the need for complex and relatively slow AFM-based nano-manipulation \cite{ribeiro-palauTwistableElectronicsDynamically2018, barabasMechanicallyReconfigurableVan2023} or constraints on device structures imposed by delamination-based disassembly \cite{taoReconfigurableElectronicsDisassembling2021}. Moreover, our process has been successful with all combinations of van der Waals materials we have tested without regard to thickness, air-sensitivity, or strain-sensitivity -- allowing it to be incorporated immediately into any heterostructure fabrication pipeline. Beyond disassembly, we note that the approach can also be used to improve the yield of conventional dry transfer assembly. For instance, the minimal friction at vdW interfaces that we leverage for sliding disassembly can also cause inadvertent misalignment between layers during assembly. By directly adjusting the misaligned layer(s), we can restore the heterostructure to the desired geometry; an example of this procedure is shown in Fig. SI 5. We envision that improving our understanding of the dynamic interfacial interactions at play would not only help eliminate existing failure modes, but could also unlock even finer control. Now equipped with a method for deterministic disassembly, vdW heterostructures should be viewed as dynamic objects that can be reconfigured almost arbitrarily, making possible a new suite of device geometries and experiments that promise to extend our understanding of 2D electron systems. 

\section{Materials and Methods}

\subsection{Preparation of manipulation slides}

\noindent We use an aperture wafer (Swiss Jewel) as a mold to prepare the PDMS pillar-on-dome structure that we use as a manipulation slide. In previous work, etched silicon \cite{gadelhaTwistedBilayerGraphene2021} or fine needles \cite{wakafuji3DManipulation2D2020} were used to create similar structures. First, we place the aperture wafer on a glass slide with the large opening facing up. We place a drop of mixed PDMS (SYLGARD 184) into the mold and cure on a hot plate at 150 $^\circ$C for 10 minutes. After cooling, we cut excess PDMS away from the mold structure. On a second glass slide, we place a small drop of PDMS and spread it into a thin layer with a razor blade onto which we flip the cured PDMS and mold structure. We again cure this at 150 $^\circ$C for 10 min. After cooling, the aperture wafer can be removed from the slide, leaving behind the microstructured stamp on glass slide, and used again.

We screen the manipulation slides prior to use to ensure the PDMS micropillar surface is flat and free from bubbles. To control the adhesion between the slide and the materials, we laminate polycarbonate (PC) film over the PDMS. Depending on the aspect ratio of the pillar, the mechanical forces from laminating the film can cause the pillar to bend, which is undesirable but can be reversed by removing the PC film. To avoid this, we tent the PC film over the PDMS -- avoiding contact between the film and PDMS pillar -- and then place the slide on a hot plate at 100 $^\circ$C – 180 $^\circ$C, which then causes the polymer to soften and uniformly coat the underlying PDMS structure.

\subsection{Manipulation of van der Waals heterostructures}

\noindent We assemble van der Waals heterostructures using PC as the transfer polymer. After releasing onto the Si/\ce{SiO2} substrate, the PC film is dissolved in chloroform, and the chip is rinsed with isopropyl alcohol before drying with nitrogen. We perform a short plasma cleaning step using either \ce{O2} or ambient air as the gas and lasting 15 s to 30 s to remove PC residues and improve adhesion between the slide and sample. The sample and slide are placed in a vdW heterostructure ``stacker" (HQ Graphene) for subsequent manipulation. The stage is heated to 70 $^\circ$C (although we note we have successfully used other stage temperatures) and the sample is aligned such that the sliding direction aligns with either the $x-$ or $y-$axis of the system. The slide is brought into contact with the sample using micropositioners. Once in contact, XY positioners are used to move the slide relative to the stage. We have successfully removed layers using stages with manual positioners, piezoelectric positioners, and stepper motor positioners. Motorized sliding has the advantage of preventing excessive out-of-plane motion compared to manual positioners. 

\subsection{Raman measurements}

\noindent Raman measurements of graphene were acquired with a Horiba XploRA Plus confocal Raman microscope system. Using a 532 nm laser with 17 mW peak power, Raman spectra maps were obtained at each stage of heterostructure manipulation. We operated at 50\% power to obtain data for 2$H$-\ce{WSe2} as the top layer, and 10\% power for all other configurations. We performed mapping over approximately the same 30 $\mu$m by 16 $\mu$m region with an 0.5 $\mu$m grid spacing along both directions. Supplemental Video 1 shows a compilation of the manipulation stages in sequence. 

Lorentzian lineshapes were fit to the observed G and 2D peaks using SciPy's \verb|optimize.curvefit|, with first-guess peak locations provided by peak-finding with SciPy's \verb|signal.find_peaks|. In our analysis of Raman peak shifts as a function of various encapsulating environments (Fig. 2f), we compute $\kappa_\mathrm{eff}$ from
\begin{equation}
\kappa_\mathrm{eff} = \frac{\sqrt{\kappa^\parallel_\mathrm{t} \kappa^\perp_\mathrm{t}} + \sqrt{\kappa^\parallel_\mathrm{b} \kappa^\perp_\mathrm{b}}}{2}
\end{equation}

\noindent where $\kappa^{\parallel (\perp)}_\mathrm{t}$ and $\kappa^{\parallel (\perp)}_\mathrm{b}$ are the dielectric constants of the top and bottom encapsulating materials for electric fields oriented parallel (perpendicular) to the interface, respectively \cite{larentisLargeEffectiveMass2018}. We use the following optical frequency dielectric constants for air, hBN, and 2$H$-\ce{WSe2}:  $(\kappa^\parallel_\mathrm{air}, \kappa^\perp_\mathrm{air}) = (1, 1)$, $(\kappa^\parallel_\mathrm{hBN}, \kappa^\perp_\mathrm{hBN}) = (4.98, 3.03)$, and $(\kappa^\parallel_\mathrm{\ce{WSe2}}, \kappa^\perp_\mathrm{\ce{WSe2}}) = (15.6, 7.7)$, respectively \cite{loudonRamanEffectCrystals2001,laturiaDielectricPropertiesHexagonal2018}.

\subsection{Sample fabrication of monolayer $H$-\ce{WSe2} devices}

\noindent The monolayer $H$-\ce{WSe2} device is assembled following the procedure described in \cite{packChargetransferContactsMeasurement2024}, but a top gate is omitted and the order of the contact structure is reversed – from top to bottom, $\alpha$-\ce{RuCl3}, $H$-\ce{WSe2}, and graphite – to move the contact gate to the bottom side of the device. After the standard sample fabrication, we use a \ce{SF6} plasma to etch a trench around the contacts in the device to separate a stationary and mobile section of the top hBN. This plasma etches hBN substantially faster than $\alpha$-\ce{RuCl3} which makes it possible to stop on the $\alpha$-\ce{RuCl3} layer without damaging the underlying contacts. The removal of the top hBN for this device is shown in Supplemental Video 2.

\subsection{Sample fabrication for 1$T'$-\ce{WTe2} device}

\noindent Working in an argon-filled glovebox, a thick hBN flake is used pick up in sequence a graphite contact, a monolayer of 1$T'$-\ce{WTe2}, a hBN bottom dielectric, and a graphite gate. The vdW stack is then deposited onto a Si/\ce{SiO2} wafer with pre-evaporated Cr/Au contacts that are aligned to the graphite electrodes. After assembly, the top hBN is removed by sliding laterally off the rest of the stack, and the device is transferred immediately to the UHV STM chamber using an inert-gas suitcase. Supplemental Video 3 shows sliding of the top hBN to reveal monolayer 1$T'$-\ce{WTe2}. We do not perform any additional processing in the STM chamber. 

\subsection{Scanning tunneling microscopy measurements}

\noindent STM and STS measurements on $H$-\ce{WSe2} were performed in a UHV chamber at room-temperature ($T$ = 298 K). The electrochemically etched tungsten tips were prepared via field emission within this chamber. Differential conductance d$I$/d$V$ spectroscopy measurements were performed using a lock-in amplifier with a 30 mV bias modulation applied to the sample, while measuring the current drained through the tip. During these measurements, gate bias voltages $V_\mathrm{g}$ and $V_\mathrm{c}$ were correlated with the tip-sample bias $V_\mathrm{b}$ to maintain fixed voltage differences. STM and STS data from the twisted bilayer graphene device were collected in a low temperature UHV STM system ($T$ = 7.6 K) with similar tip preparation and measurement methods.

STM and STS measurements of 1$T'$-\ce{WTe2} were conducted in a Scienta Omicron Polar system with a base temperature $T \approx 5$ K. The STM tips were made from chemically etched tungsten wires, which were further conditioned on a Cu(111) single crystal through a combination of pulsing and poking, such that poke marks were localized and the Cu(111) surface state was present \cite{liuVisualizingBrokenSymmetry2022,crommieConfinementElectronsQuantum1993}. To preserve the tip apex, each conditioned tip was engaged directly onto 1$T'$-\ce{WTe2} via capacitance-guided navigation \cite{liSelfnavigationScanningTunneling2011}, and were further conditioned on the Au electrodes as needed. During the measurements, a voltage bias -$V_\mathrm{b}$ was applied to the tip with respect to the grounded sample, and the tunneling current $I$ was collected downstream from the tip. The measured topography was FFT-filtered to remove frequency components associated with vibrational noise.

The differential conductance d$I$/d$V$ signal was measured with standard lock-in techniques using a peak-to-peak excitation of 5 mV at 419 Hz. Gate-dependent spectroscopy measurements were performed by first stabilizing the tip at a set $V_\mathrm{g}$, $V_\mathrm{b}$, and $I$, then opening the feedback loop and sweeping $V_\mathrm{g}$ to the desired value, followed by sweeping $V_\mathrm{b}$ while recording d$I$/d$V$. This ensures a constant tip-sample separation during the measurement.

\subsection{Scanning near-field optical microscopy measurements}

\noindent Scanning near-field optical microscopy (SNOM) images of moir\'e domains were acquired with a commercial phase-resolved scattering type scanning optical microscope system (Neaspec) operated under continuous-wave mid-infrared excitation from a quantum cascade laser (Hedgehog by Daylight Solutions) tuned to wavelength $\lambda = $ 9.6 $\mu$m. The laser was focused onto the apex of an AFM tip oscillated in tapping mode. The scattered near-field signal was collected by a cryogenic HgCdTe detector (Kolmar Technologies). The complex near-field signals (amplitude and phase) were retrieved by demodulating at higher harmonics of the tip-tapping frequency using a pseudo-heterodyne interferometric scheme, wherein the scattered field is interfered with a modulated reference arm at the detector to suppress far-field background. The images shown in the manuscript correspond to the phase channel demodulated at the fourth harmonic.

\subsection{Lateral Force Microscopy Measurements}

\noindent Lateral force microscopy (LFM) was measured with an Asylum Jupiter XR AFM using an AC240 cantilever. We estimate that approximately 40 nN of normal force was applied during the LFM measurements.

\section{Acknowledgments}

\noindent We acknowledge technical assistance from Harvey Runyon and Aidan Gregerson, and fruitful discussions with Rebeca Ribeiro-Palau. \textbf{Funding:} This research is primarily supported by Programmable Quantum Materials, an Energy Frontier Research Center funded by the US Department of Energy (DOE), Office of Science, Basic Energy Sciences (BES), under award DE-SC0019443, and also in part by the Gordon and Betty Moore Foundation through Grant GBMF12766 to A.D. (instrumentation). C.R.D. acknowledges support by a Brown Investigator Award, a program of the Brown Institute for Basic Sciences at the California Institute of Technology. A portion of this work made use of shared fabrication facilities at Columbia provided through the NSF MRSEC DMR-2011738. A portion of this work made use of shared fabrication facilities at UW by NSF MRSEC DMR-2308979. Synthesis of hBN (K.W., T.T.) was supported by the JSPS KAKENHI (Grant Numbers 21H05233 and 23H02052), the CREST (JPMJCR24A5), JST and World Premier International Research Center Initiative (WPI), MEXT, Japan.

\section{Author contributions}

\noindent K.V.F., A.D., J.P., and C.R.D. developed the sliding disassembly process. J.P., K.V.F., S.G., K.T.C., and F.M. fabricated devices. K.V.F. performed and analyzed Raman measurements. J.P. and S.G. performed and analyzed electrical transport measurements. J.P. and C.G. performed and analyzed AFM topography, and LFM moir\'e characterization. K.T.C., F.M., E.T., and M.Y. performed and analyzed STM measurements of 1$T'$-\ce{WTe2} devices. S.G., X.W., and A.N.P. performed and analyzed STM measurements of $H$-\ce{WSe2} and graphene devices. J.P., Z.Z., and D.N.B. conducted and analyzed SNOM measurements. K.W. and T.T. provided hBN crystals. J.P., K.V.F., C.R.D., and A.D. wrote the manuscript with input from all authors. C.R.D. and A.D. supervised the project.

\section{Competing interests} \noindent The authors declare no competing interests.

\newpage
\renewcommand{\figurename}{Fig. SI}
\setcounter{figure}{0}

\begin{figure*}
    \makebox[\textwidth][c]{\includegraphics[scale=1.0]{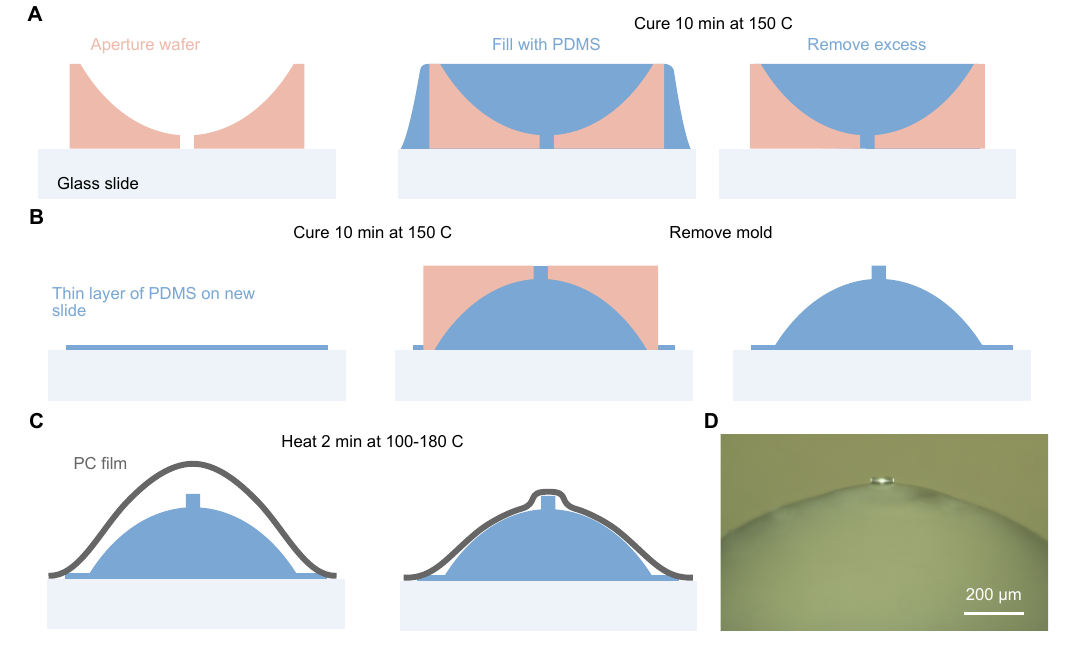}}
\caption{\textbf{Process for preparing manipulation slides.} \textbf{(A)} An aperture wafer used as a mold is filled with PDMS and cured. After curing, excess PDMS outside the mold is removed. \textbf{(B)} On a new slide, a thin layer of PDMS is spread to adhere to the molded PDMS. The mold and PDMS structure is flipped onto this slide and cured. Finally, the mold is removed. \textbf{(C)} A PC film is tented over the slide to prevent bending the PDMS pillar. By heating at 100-180 $^\circ$C for 2 min, the film conforms to the PDMS. \textbf{(D)} Profile optical image of a representative PDMS dome and micropillar.}
\end{figure*}

\begin{figure*}
    \makebox[\textwidth][c]{\includegraphics[scale=1.0]{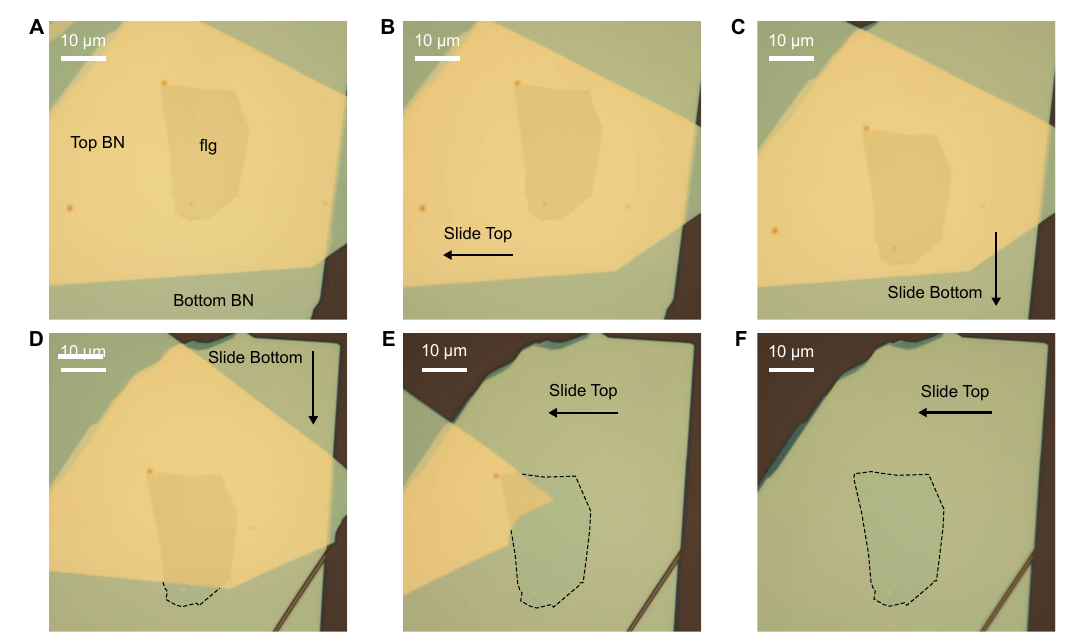}}
\caption{\textbf{Manipulation of top and bottom layers.} \textbf{(A)} Stack of few layer graphite (flg) encapsulated in hBN on Si/\ce{SiO2}. \textbf{(B)} An image after partially sliding the top hBN to the left, \textbf{(C),(D)} after sliding the bottom hBN down in two steps, and \textbf{(E), (F)} and after sliding the top hBN to fully expose the flg in two steps.}
\end{figure*}

\begin{figure*}
    \makebox[\textwidth][c]{\includegraphics[scale=1.0]{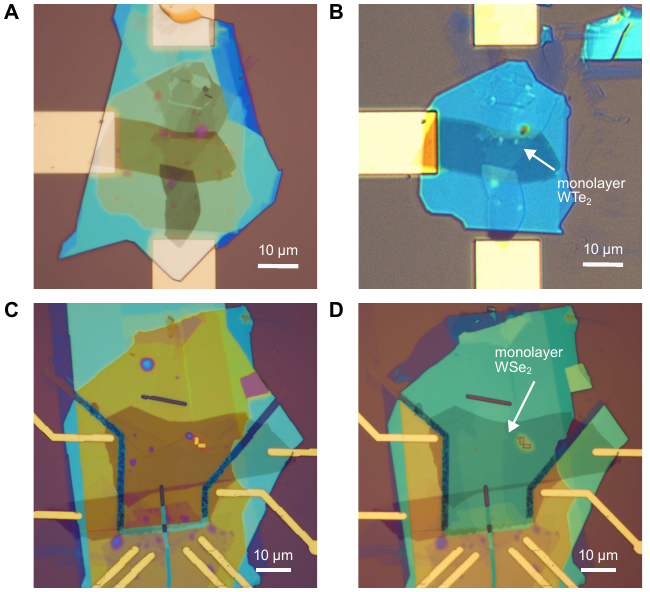}}
\caption{\textbf{Optical images of STM devices.} \textbf{(A)} Images of the monolayer 1$T'$-\ce{WTe2} sample before and \textbf{(B)} after sliding the top hBN layer. \textbf{(C)} Optical image of the monolayer $H$-\ce{WSe2} device before and \textbf{(D)} after removing the region of hBN covering the device.}
\end{figure*}

\begin{figure*}
    \makebox[\textwidth][c]{\includegraphics[scale=1.0]{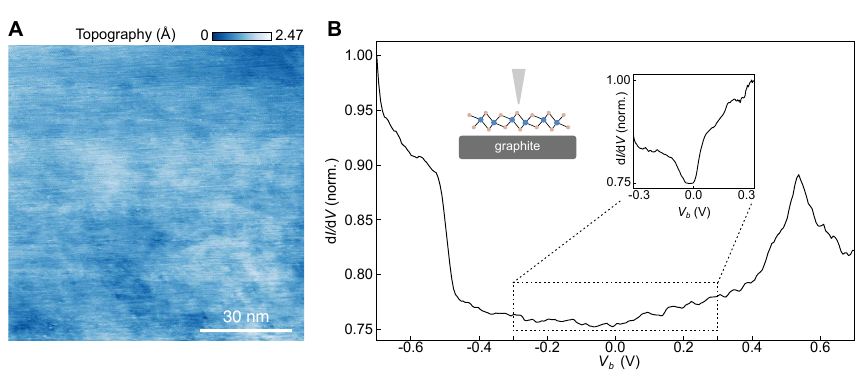}}
\caption{\textbf{Additional STM of 1$T'$-\ce{WTe2}.} \textbf{(A)} Large area topograph of gated 1$T'$-\ce{WTe2} device. \textbf{(B)} d$I$/d$V$ spectroscopy of monolayer 1$T'$-\ce{WTe2} on graphite. (top left) Cartoon schematic of the sample and STM tip. (top right) Zoomed-in d$I$/d$V$ spectrum at smaller bias.}
\end{figure*}

\begin{figure*}
    \makebox[\textwidth][c]{\includegraphics[scale=1.0]{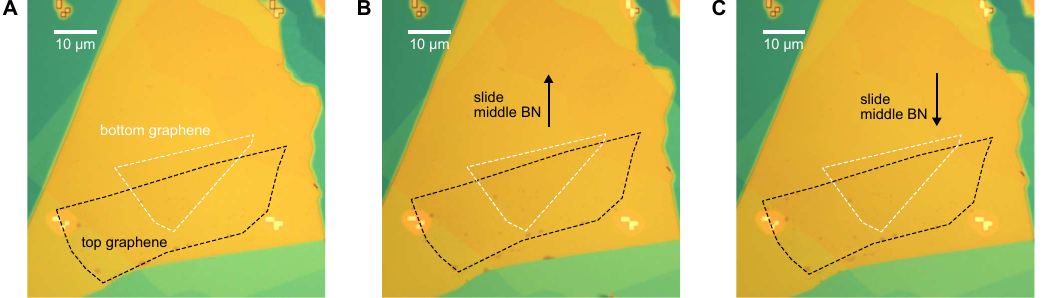}}
\caption{\textbf{Adjusting layer alignment.} \textbf{(A)} Image of heterostructure with two layers of graphene separated by a layer of hBN. The boundaries of the two graphene layers are outlined. \textbf{(B)} Image of the same heterostructure after sliding the middle hBN upwards to change the relative alignment between the top and bottom graphene layers. \textbf{(C)} Optical image after sliding the middle BN downwards slightly to further adjust the relative positions.}
\end{figure*}

\end{document}